\documentclass[12pt]{article}

\usepackage{amsmath}
\usepackage{graphicx}
\usepackage{hyperref}
\usepackage[FIGTOPCAP]{subfigure}

\begin{document}

\title{How does a heavy symmetric top rise?}
\author{        V. Tanr{\i}verdi \\
tanriverdivedat@googlemail.com \\
Address: Bahad{\i}n Kasabas{\i} 66710 Sorgun-Yozgat TURKEY
}

\date{}

\maketitle

\begin{abstract}
	The rise of the top is studied by considering slipping friction.
	For this case, equations of motion are obtained by using Euler equations for a heavy symmetric top with a hemispherical peg.
	Different situations are considered to see how the rise takes place, which can help understand the rise of the top.
\end{abstract}

\section{Introduction}

The rise of the top is one of the impressive physical phenomena for rigid body rotations.
The explanation of the rise comes from the friction \cite{Gray}, and this has been known for more than a century. 
There are some books on classical mechanics and rigid body rotations considering the rise of the top \cite{BargerOlson, Perry, Jellett}, however, they are not easily understandable by students.
On the other hand, the rise of the top is a useful example of how to include torque in rigid body rotations and can be very helpful in education.

In previous works, the rise of the top is considered in the body reference frame, whose origin is at the center of mass.
In this type of consideration, one needs to consider torque originating from the reaction force of the surface \cite{Jellett, Moffatt, Parkyn, Yogi} that makes the situation more complicated.

We will consider the problem in the body reference frame whose origin is at the radial center of the spherical peg.
This gives simpler resultant equations and makes the effect of the friction easily understandable.
In the previous work, we have compared Jellet's model and the model applied in this work \cite{Tanriverdi_rise}.
In previous version (version 3) of that work, we have explained the rise of the top, though, in the latest version, this explanation is not available.
In this work, we will modify and explain the rise of the top in detail, which can be helpful to students.

In this work, we will consider the problem by using Euler equations; usage of them is necessary since friction, a non-conservative force, is considered.
In section \ref{two}, we will obtain equations of motion.
In section \ref{three}, we will numerically solve different situations to see how the rise occurs.
Then, we will conclude in section \ref{four}.

\section{Rise of the top}
\label{two}

\begin{figure}[!h]
        \begin{center}
                \subfigure[]{\includegraphics[width=5.5cm]{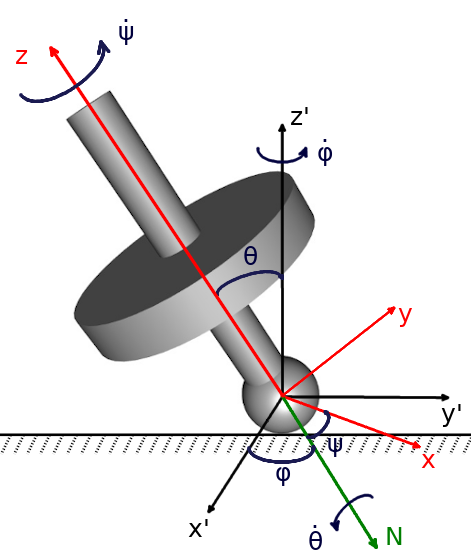}}
                \subfigure[]{\includegraphics[width=5.5cm]{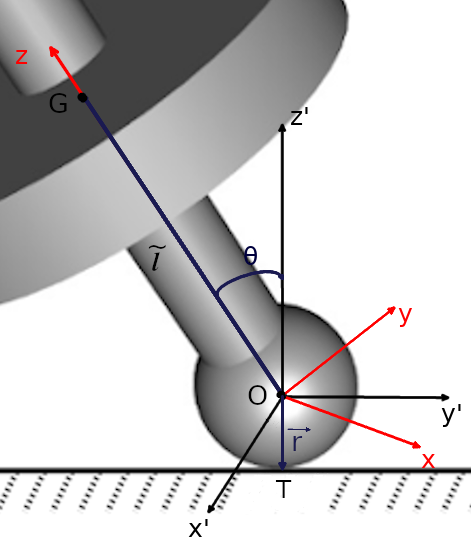}}
                \caption{ a) Heavy symmetric top with a spherical peg, stationary reference frame ($x',y',z'$), center of peg-body reference frame ($x,y,z$), line of nodes $N$, Euler angles ($\theta, \phi, \psi$), angular velocities ($\dot \theta, \dot \phi, \dot \psi$). b) Peg of the top, center of peg $O$, touchpoint $T$, center of mass $G$, $\vec r$ and $\tilde l$.
                }
        \label{fig:hst}
        \end{center}
\end{figure}

Let us consider a symmetric top having a spherical peg with radius $R$, shown in figure \ref{fig:hst}.
And, consider that its mass is $M$, the distance between the center of the peg and the center of mass is $\tilde l$ and moments of inertia $I_y=I_x$ and $I_z$ in the body reference frame whose origin is the peg's radial center.
Due to rotations, there are some changes in the reaction force of the surface.
However, they are small since $\dot \theta$ is small, and we will ignore them and say $N=M g$.
The touchpoint $T$ of the top has a velocity with respect to the peg's center, and it can be found by using $\vec v=\vec w \times \vec r$
where $\vec w$ is the top's angular velocity and $\vec r$ is the vector from the peg's center to the touchpoint.

Since friction has to be included to obtain the rise of the top, it is better to use Euler equations which can be written for a symmetric top ($I_y=I_x$) as 
\begin{eqnarray}
       \tau_x&=&I_x \dot w_x+w_y w_z (I_z-I_x),  \nonumber \\
       \tau_y&=&I_x \dot w_y+w_x w_z (I_x-I_z), \label{eeq} \\
       \tau_z&=&I_z \dot w_z, \nonumber
\end{eqnarray}
where $\vec \tau$ is torque.
These equations are in the body reference frame.

One can write angular velocities in terms of Euler angles as
\begin{eqnarray}
        w_x&=&\dot \theta \cos \psi+\dot \phi \sin \theta \sin \psi, \nonumber \\
        w_y&=&-\dot \theta \sin \psi+ \dot \phi \sin \theta \cos \psi, \label{ang_vel}\\
        w_z&=&\dot \psi+\dot \phi \cos \theta, \nonumber 
\end{eqnarray}
where $\theta$ is the angle between the stationary $z'$-axis and body $z$-axis and describes rotation around the line of nodes, 
$\phi$ is the angle describing rotations around the stationary $z'$-axis, 
and $\psi$ is the angle describing rotations around the body $z$-axis. 
These can be seen in figure \ref{fig:hst}.

By using $\vec r=-R \hat z'$ and $\hat z'= \sin \theta \sin \psi \hat x+\sin \theta \cos \psi \hat y+ \cos \theta \hat z$, we can find velocity of the toucpoint with respect to the peg's center in terms of Euler angles as
\begin{eqnarray}
	\vec v=R &[&(\dot \theta \cos \theta \sin \psi+\dot \psi \sin \theta \cos \psi)\hat x+(\dot \theta \cos \theta \cos \psi-\dot \psi \sin \theta \sin \psi)\hat y \nonumber \\
        & &+(-\dot \theta \sin \theta)\hat z].
        \label{slipping_velocity}
\end{eqnarray}
The magnitude of this velocity can be obtained as $|\vec v|=R \sqrt{\dot \theta^2 +\dot \psi^2 \sin^2 \theta}$.

In daily life, the top can roll on the surface.
However, considering rolling makes the situation complicated and the aim of this work is only to show how the top rises by friction.
Then, we will ignore the rolling motion and consider that the top slips on the surface.
Hence, friction force can be written as $\vec f=- k N \vec v / |\vec v|$, where $k$ is a positive friction constant.

Now, we can obtain torque due to friction force by using $\vec \tau=\vec r \times \vec f$ as
\begin{eqnarray}
        \vec \tau= &\frac{k M g R^2} {|\vec v|}&  [(-\dot \theta \cos \psi+\dot \psi \sin \theta \cos \theta \sin \psi)\hat x\nonumber \\
        & & \,+(\dot \theta \sin \psi+\dot \psi \sin \theta \cot \theta \cos \psi)\hat y+(-\dot \psi \sin^2 \theta)\hat z],
	\label{velocityT}
\end{eqnarray}
and the gravitational torque can be obtained as
\begin{equation}
        \vec \tau_g=-Mg \tilde l \sin \theta (-\cos \psi \hat x+\sin \psi \hat y).
\end{equation}

As it is seen from equation \ref{velocityT} velocity (and dependently friction) does not depend on $\dot \phi$, which is the natural result of considering the touchpoint as a point.
On the other hand, they depend on the other two angular velocities $\dot \theta$ and $\dot \psi$, and in general $|\dot \psi| >> |\dot \theta|$.
By considering this, ignoring dissipation due to motion related to $\theta$ is a plausible assumption. 
Then, by ignoring it, we can write components of torque as
\begin{eqnarray}
        \tau_x&=& Mg \tilde l \sin \theta \cos \psi + \frac{k M g R^2 (\dot \psi \sin \theta \cos \theta \sin \psi ) }{|\vec v|}, \nonumber \\ 
	\tau_y&=& Mg \tilde l \sin \theta \sin \psi + \frac{k M g R^2 (\dot \psi \sin \theta \cos \theta \cos \psi ) }{|\vec v|} \\
        \tau_z&=&- \frac{k M g R^2 \dot \psi \sin^2 \theta }{|\vec v|}. \nonumber 
\end{eqnarray}

Now, we can include these components of torque in Euler equations and obtain the following equations with some algebra as
\begin{eqnarray}
       \ddot \theta&=& -\frac{I_z \dot \phi \sin \theta}{I_x}(\dot \psi+\dot \phi \cos \theta)+\dot \phi^2 \sin \theta \cos \theta+ \frac{Mg \tilde l }{I_x} \sin \theta,  \nonumber \\
       \ddot \phi&=& \frac{I_z \dot \theta}{I_x \sin \theta}( \dot \psi +\dot \phi \cos \theta)- \frac{ 2 \dot \theta \dot \phi \cos \theta}{ \sin \theta} +\frac{k M g R^2 \dot \psi \cos \theta}{I_x |\vec v|},  \label{diffeqns_odp} \\
       \ddot \psi&=& - \frac{I_z \dot \theta \cos \theta}{I_x \sin \theta} ( \dot \psi +  \dot \phi \cos \theta )+\frac{2 \dot \theta \dot \phi \cos^2 \theta}{\sin \theta} +\dot \theta \dot \phi \sin \theta \nonumber \\
	& & -\frac{k M g R^2 \dot \psi }{|\vec v|}\left(\frac{\cos^2 \theta}{I_x}+\frac{\sin^2 \theta}{I_z}\right). \nonumber
\end{eqnarray}
As it is seen, the considered dissipation shows itself in angular accelerations $\ddot \phi$ and $\ddot \psi$.

These three equations of motion are coupled equations, and any change in one of these will affect the other one.
Hence, it can be very difficult to comment on these; however, one can still consider some aspects.
And in here, we will consider $\ddot \theta$ a bit further since we are considering the rise of the top.
$\ddot \theta$ is a function of $\theta$, $\dot \psi$ and $\dot \phi$, and it depends quadratically on $\dot \phi$.
If one considers fixed $\theta$ and $\dot \psi$ values, then there are, in general, two roots giving regular precession.
Due to $\ddot \theta$'s quadratic nature, $\ddot \theta$ should be either positive or negative when $\dot \phi$ value is between these two roots.
For ordinary tops, $\theta$ is smaller than $\pi/2$.
And, when $I_x>I_z$, $\ddot \theta$ becomes negative between these two roots for ordinary tops, and vice versa.
When $\ddot \theta$ is negative, the top rises.

The dissipative term in $\ddot \phi$, $k M g R^2 \dot \psi \cos \theta/I_x |\vec v|$, has a positive sign and that term increases the magnitude of the $\dot \phi$.
For the motion of an ordinary top, if $\dot \phi$ is smaller than the smaller root giving regular precession, then an increase of $\dot \phi$ results in the rise of the top provided that $\theta<\pi/2$ and $I_z<I_x$ \cite{Tanriverdi_wdwuc}.

\section{Numerical solution}
\label{three}

We will numerically solve equations \eqref{diffeqns_odp} to see the rise of the top.
We will consider a top with $I_x=8.52 \times 10^{-5}  \,kg\,m^2$, $I_z=7.25 \times 10^{-5}  \,kg\,m^2$, $M=110 \,gr$, $\tilde l=20 \,mm$ and $R=7 \,mm$.
Initial values will be taken as $\theta_0=0$, $\dot \theta_0=0$, $\dot \phi_0=1.75 \,rad\,s^{-1}$ and $\dot \psi_0=170 \,rad\,s^{-1}$ ( $\phi_0=0$ and $\psi_0=0$ ).
This configuration is very close to regular precession, and it is considered to reduce nutation, which is not necessary to obtain the top's rise.
Some of the terms in equations \eqref{diffeqns_odp} go to infinity as $\theta$ goes to zero, and to avoid these infinities, the numerical solution is cut when the top comes to the nearly upright position, i.e. $\theta=0.02 \,rad$. 

\begin{figure}[h!]
        \begin{center}
        \subfigure[$\theta$]{
                \includegraphics[width=4.2cm]{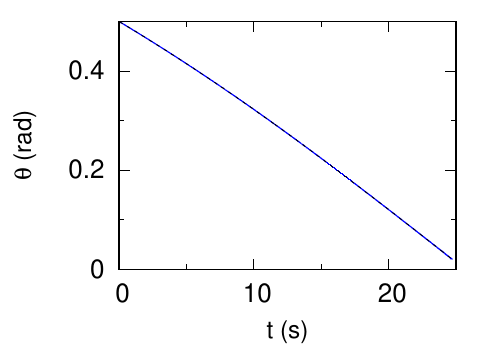}
                }
        \subfigure[$\dot \theta$]{
                \includegraphics[width=4.2cm]{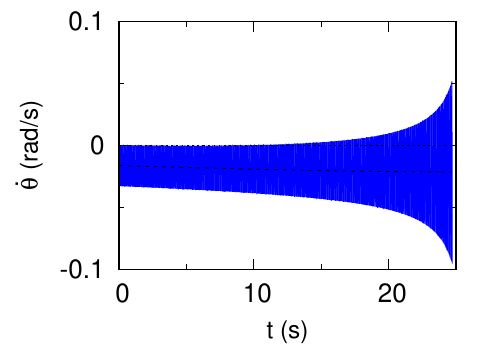}
                }

        \subfigure[$\dot \phi$]{
                \includegraphics[width=4.2cm]{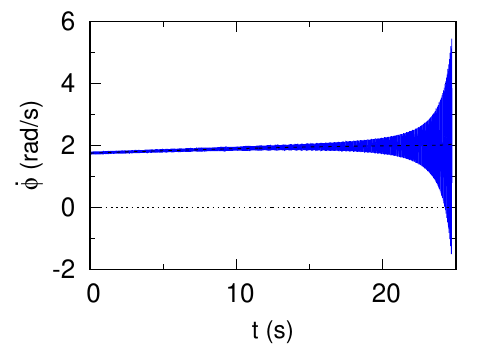}
                }
        \subfigure[$\dot \psi$]{
                \includegraphics[width=4.2cm]{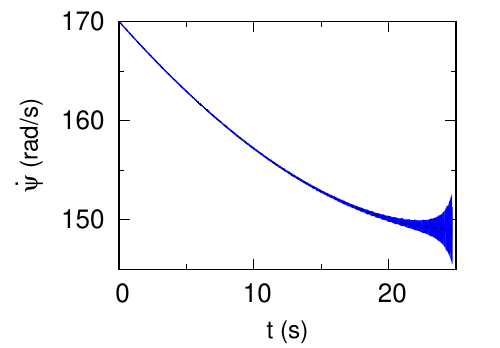}
                }
		\caption{Results of the numerical solution for full equations \eqref{diffeqns_odp} for $\theta$ (a), $\dot \theta$ (b), $\dot \phi$ (c) and $\dot \psi$ (d). Continuous (blue) curves show results of numerical solution, which have a bulk structure in some places due to nutation, dashed (black) curves show nutation average. Initial values: $\theta_0=0.5\,rad$, $\dot \theta_0=0$, $\dot \phi_0=1.75 \,rad\,s^{-1}$ and $\dot \psi_0=170 \,rad\,s^{-1}$. 
                }
        \label{fig:thetaphipsi_d}
        \end{center}
\end{figure}

\begin{figure}[h!]
        \begin{center}
        \subfigure[]{\includegraphics[width=4.2cm]{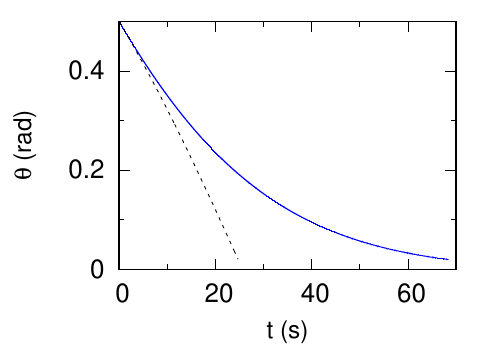}}
        \subfigure[]{\includegraphics[width=4.2cm]{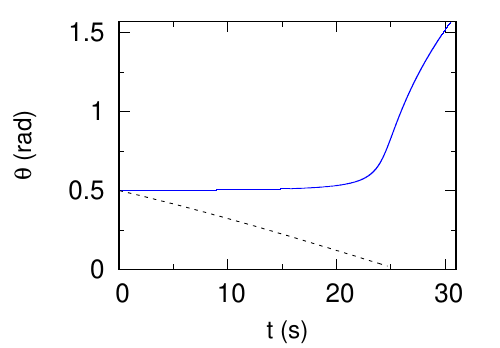}}
                \caption{ Results of the numerical solution for inclination angle $\theta$ when different dissipative effects are considered.
		Continuous (blue) lines show results of numerical solution for: (a) equations \ref{diffeqns_odp} without the dissipative term in $\ddot \psi$, (b) equations \ref{diffeqns_odp} without the dissipative term in $\ddot \phi$.
		Dashed lines (black) show nutation average results for full equations \ref{diffeqns_odp}.
                }
        \label{fig:thetaphipsi_e}
        \end{center}
\end{figure}

Results of the numerical solutions for equations \eqref{diffeqns_odp} for $\theta$ can be seen in figure \ref{fig:thetaphipsi_d}.
One can see from figure \ref{fig:thetaphipsi_e}(a) that the top rises to the nearly upright position in around $25$ seconds.
There are some small nutations, figure \ref{fig:thetaphipsi_e}(b), which do not affect the general motion.
It can be seen that the average of $\dot \theta$ is negative as required for the rise.
From figure \ref{fig:thetaphipsi_e}(c), it can be seen that the average of $\dot \phi$ slightly increases, which is the main reason for the rise.
$\dot \phi$ takes negative values at the end, which means that the top makes the looping motion at the later stages of motion \cite{Goldstein}.
The average value of $\dot \psi$ decreases as a result of dissipation, figure \ref{fig:thetaphipsi_e}(d).
In real situations, after the rise of the top, dissipation slows down the top, and then it falls.

In the numerical solution, there is some increase in fluctuations of angular velocities at the end, and such fluctuations are not seen in $\theta$, see figure \ref{fig:thetaphipsi_e}(a).
In fact, the observed fluctuations are the results of very small fluctuations in small $\theta$ values.
In real situations, such fluctuations, in general, are damped by air dissipation which we did not consider in this work.

\begin{figure}[h!]
        \begin{center}
                \subfigure[]{\includegraphics[width=4.2cm]{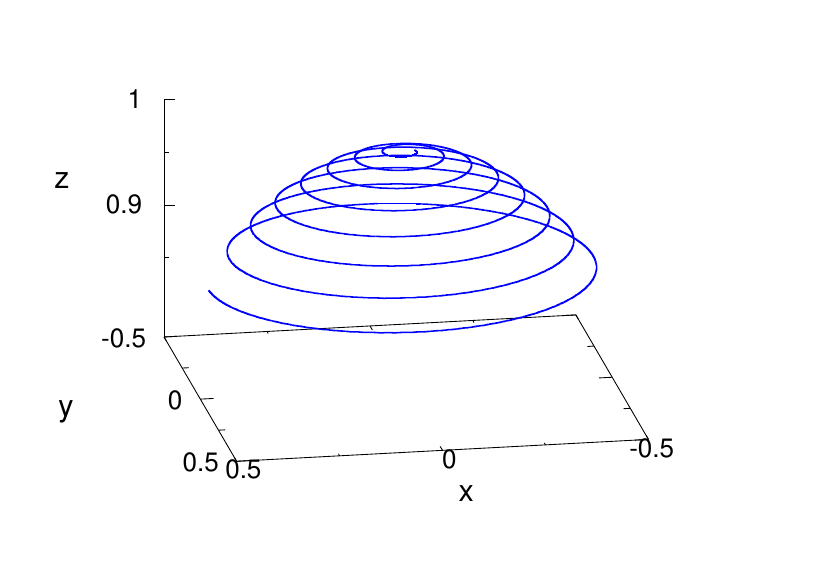}}
                \subfigure[]{\includegraphics[width=4.2cm]{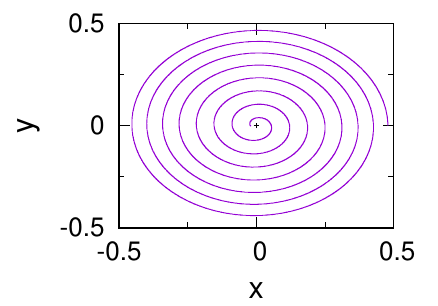}}
                \caption{Shapes for the locus (a) and its projection on to $xy$-plane (b) for the rise of the top.
                Initial values are given in figure \ref{fig:thetaphipsi_d}, and the solution is obtained by considering full equations \eqref{diffeqns_odp}.
		Animated version is available at \href{https://youtu.be/gcAnamQHmBM}{https://youtu.be/gcAnamQHmBM}.
                }
        \label{fig:tt_prj_f}
        \end{center}
\end{figure}

One can see the three-dimensional plot of the change of symmetry axis's position, shapes for the locus, and its projection in figure \ref{fig:tt_prj_f}. 
It can be seen that the top rises slowly with a spiraling structure, which is different than spiralling motion \cite{Routh, Tanriverdi_abequal}.
We previously mentioned that $\dot \phi$ has negative values at the end of the solution, implying that the top makes the looping motion.
Such a motion is observed; however, it is not resolved in the plot due to the very small structure.

Now, we will consider and numerically solve two more cases to clearly see the reason of the rise: Equations \eqref{diffeqns_odp} without the dissipative term in $\ddot \psi$ and equations \eqref{diffeqns_odp} without the dissipative term in $\ddot \phi$.
We will take the parameters of the top and initial values the same as the numerically solved example.
Results of the numerical solution for $\theta$ can be seen in figure \ref{fig:thetaphipsi_e}(a) when only the dissipative term in $\ddot \phi$ is included, and the rise of the top is observed.
This shows that the rise is related to the dissipative term in $\ddot \phi$.
We should note that the rise to the nearly upright position takes some longer time.
This occurs due to not including the dissipative term in $\ddot \psi$: $\dot \psi$ does not decrease, and $\dot \phi$ requires bigger values for the rise.

In the other case, we will set the dissipative term in $\ddot \phi$ to zero and keep the dissipative term in $\ddot \psi$.
Results of the numerical solution for $\theta$ can be seen in figure \ref{fig:thetaphipsi_e}(b).
It can be seen from that figure that the top does not rise and fall, which clearly shows that the dissipative term in $\ddot \phi$ is the cause of the rise, and the dissipative term in $\ddot \psi$ does not directly contribute to the rise of the top.
Results of the numerical solution show that $\dot \psi$ decreases quickly, and the reason for this quick decrease is the absence of the dissipative term in $\ddot \phi$.
Around $t \approx 24 \,s$, there is a fast fall of the top because $|a|=I_z |\dot \psi+\dot \phi \cos \theta|/I_x$ takes smaller values than $\sqrt{4 Mg \tilde l/I_x}$ which corresponds to "weak top" condition \cite{KleinSommerfeld}.

\section{Conclusion}
\label{four}

The rise of the top with a hemispherical peg is studied.
We have considered the problem in the body reference frame, whose origin is at the peg's radial center.
Results show that rise is provided by the term $\frac{k N R^2 \dot \psi \cos \theta}{I_x |\vec v|}$ in $\ddot \phi$ which is the result of the friction at the touchpoint.
We have also seen from the results that if the dissipation in $\ddot \psi$ is removed, the rise time becomes longer.
Then, we can say that in real situations, due to the presence of air dissipation $\dot \psi$ decreases and the top rises faster.
In previous works, this problem is studied in the body reference frame whose origin is at the center of mass.
If the origin is taken at the center of mass, the rise term depends on $\tilde l$ also, which is different from the studied model.

The rise of the top deserves a more detailed explanation which holds for both models.
We have seen that the rise of the top is directly related to the term in $\ddot \phi$ originating from the slipping friction at the touchpoint due to the rotation related to the spin angular velocity as it is known previously.
This term causes an increase in the magnitude of $\dot \phi$, and an increase of $\dot \phi$ can result in negative $\ddot \theta$.
This is stated as "Hurry on the precession, and the body rises in opposition to gravity." by Perry.
We should note that this happens if the precession angular velocity is smaller than the smaller root for the regular precession for tops satisfying $\theta<\pi/2$ and $I_x>I_z$.
Then, as a summary, one can say that \textit{the slipping friction at the touchpoint due to rotation related to the spin angular velocity generates a term in the precession angular acceleration, which increases the precession angular velocity, and this increase results in the rise of the top by making the angular acceleration for inclination angle negative.}

We should note that for the rise of the top, the increase of $\dot \phi$ is not a necessary condition.
If dissipation reduces the magnitude of $\dot \psi$ more quickly, then one can observe the rise of the top without any increase of $\dot \phi$.
This can be understood by studying $\ddot \theta$ considering different situations.

\end{document}